\documentclass[cmbright,doublespace]{mmaauth}%For paper submission

\usepackage{moreverb, amssymb,amsmath}

\usepackage[dvips,colorlinks,bookmarksopen,bookmarksnumbered,citecolor=red,urlcolor=red]{hyperref}

\newcommand\BibTeX{{\rmfamily B\kern-.05em \textsc{i\kern-.025em b}\kern-.08em
T\kern-.1667em\lower.7ex\hbox{E}\kern-.125emX}}

\def\volumeyear{2010}
\newtheorem{theorem}{Theorem}
\newtheorem{lemma}{Lemma}
\newtheorem{remark}{Remark}

\def\bea{\begin{eqnarray*}}
\def\eea{\end{eqnarray*}}
\def\disp{\displaystyle}
\def\e{\varepsilon}

\def\t{\theta}

\def\vf{\varphi}
\def\g{\gamma}
\def\r{\rho}
\def\W{\Omega}
\def\bq{{\bf q}}
\def\bv{{\bf v}}
\def\bA{{\bf A}}
\def\bx{{\bf x}}
\def\bb{{\bf b}}

\def\bD{{\bf D}}
\def\bT{{\bf T}}
\def\bA{{\bf A}}
\def\bB{{\bf B}}
\def\bq{{\bf q}}
\def\bJ{{\bf J}}
\def\bS{{\bf S}}
\def\bn{{\bf n}}

\def\cP{{\mathcal P}}
\def\div{\nabla\cdot}
\def\rot{\nabla\times}

\begin{document}

\runninghead{A. Berti and I. Bochicchio}

\title{A mathematical model for phase separation: a generalized Cahn--Hilliard equation.}

\author{A.~Berti\affil{a} and I.~Bochicchio\corrauth}

\corraddr{Dipartimento di Matematica ed Informatica, Universit\'a
degli Studi di Salerno, Via Ponte Don Melillo, 84084 Fisciano
(SA), Italy. E-mail: ibochicchio@unisa.it}

\address{\affilnum{a}Faculty of Engineering, University e-Campus, 22060 Novedrate (CO), Italy. E-mail: alessia.berti@uniecampus.it
}

\begin{abstract}
In this paper we present a mathematical model to describe the
phenomenon of phase separation, which is modelled as space
regions where an order parameter changes smoothly. The model
proposed, including thermal and mixing effects, is deduced for an
incompressible fluid, so the resulting differential system couples
a generalized Cahn--Hilliard equation with the Navier--Stokes
equation. Its consistency with the second law of thermodynamics in
the classical Clausius-Duhem form is finally proved.

\end{abstract}

\MOS{74A50; 80A17}


\maketitle

%\footnotetext[2]{Please ensure that you use the most up to date
%class file,
%available from the MMA Home Page at\\
%
%\href{http://www3.interscience.wiley.com/journal/2197/home}{\texttt{www3.interscience.wiley.com/journal/2197/home}}}

\vspace{-6pt}

\section{Introduction}
The mechanism by which a mixture of two or more components separate into distinct regions
(or phases) with different chemical compositions and physical properties is usually named
{\it spinodal decomposition} or {\it phase separation}. This mechanism differs from classical nucleation in that phase separation is much more subtle, and occurs uniformly throughout the material, not just at discrete nucleation sites.

Typically the phenomenon of spinodal decomposition occurs when a mixture of two different species, say $A$ and $B$, forming a single homogeneous phase at a temperature $\theta_m$ greater than the critical temperature $\theta_0$, is rapidly cooled to a temperature where the homogeneous state is unstable. The resulting inherent instability leads
to composition fluctuations, and thus to instantaneous phase separation.

The most common experimental examples of spinodal decomposition occur in metallic alloys \cite{B-T,R-H} and glassy mixtures \cite{ABB,TMH}. For example an Al--rich Al--Zn alloy, when quenched rapidly from above 400 $^{\circ }C$ and then annealed at temperatures in the neighborhood of 100 $^{\circ }C$, is known to decompose into Al-- and Zn--rich regions via the spinodal mechanism \cite{LBM}.

The basic theory of spinodal decomposition has been developed, primarily from a metallurgic point of view by Hillert \cite{Hill}, Cahn \cite{C,C2}, Hilliard \cite{H1} and Cook \cite{Cook}. Subsequently Cahn developed a more general linearized theory of spinodal instability pointing out the essential role played by nonlinear effects in determining the nature of the instability and then in limiting its growth \cite{LBM}.

The phase separation is often described in the framework of phase--field modelling, in that the interface between the two pure phases is not sharp but is regarded as  a region of finite width having a gradual variation of different physical quantities. In addition, to distinguish one phase from the other, it is necessary to select a quantity which differs in the two phases. Since Landau, such a quantity is called {\it order parameter} and it assumes distinct values in the bulk phases away from the interfacial regions over which it varies smoothly.

Interpreting the order parameter as the concentration of one of the two metallic components of the binary alloy, Cahn and Hilliard \cite{CH,C} introduce the so-called Cahn-Hilliard equation which describes the evolution of the concentration field in a binary alloy.

In the present paper, we present a generalized mathematical model capable of describing a phase separation into the Cahn-Hillard theory. Precisely, we consider a mixture of two incompressible fluids with comparable densities but different viscosity, and we assume that our system can be described by a single scalar order parameter $c$, which we can visualize as the difference of local mass fraction (concentration) of the two components of a binary solution.
In addition, we suppose that the density of mixture does not depend on the composition of the mixture
({\it i.e.} $\rho(c)\,=\,\rho_0$) such that the general mass balance equation of the mixture degenerates to the solenoidal condition.
Following \cite{LT}, in Sect. 3, we postulate that $c$ obeys a diffusion equation.

The aim of our paper is to propose a model accounting for the fluid motion. In particular, besides the classical coupling between the Cahn--Hilliard and Navier--Stokes equations, due to the presence of the material derivative of the order parameter in the Cahn--Hilliard equation and of a surface tension source term in the Navier--Stokes equation (see \cite{gurtin,LT}), we suppose that the chemical potential may depend on the curl of the velocity of the mixture (see Sect. 4). The effect of the velocity can be interpreted as an increase of the temperature which controls the phase separation.

In Sect. 5, we modify the classical Navier--Stokes equation by adding a reactive stress,
accounting for the capillary forces due to surface tension, and a skew tensor consistent with the presence of
internal structure due to the mixture \cite{cosserat} which guarantees the coupling with the Cahn--Hilliard equation.
Finally, in Sect. 6, we prove the compatibility with thermodynamics of our model by expressing the second law in the classical Clausius--Duhem inequality.

% ==================================================================
% ==================================================================
% ==================================================================

\section{Phase--field modelling} Let $\Omega \subset \mathbb{R}^3$
be a fixed bounded domain, which is completely filled by a mixture
of two incompressible fluids $A$ and $B$, and let $\partial \W$ be
its smooth boundary with unit outward normal $\bn$. For the sake
of simplicity, we suppose that the densities $\r_A, \r_B$ of both
components as well as the density $\r_0$ of the mixture are
constant and we assume
\begin{equation}\label{constant}
\r_A=\r_B=\r_0\,=\, const.
\end{equation}
Let $m$ be the total mass of the mixture, {\it i.e.}
$$
m= \int_\W \r_0 dv
$$
and let $m_A$, $m_B$ be the masses of each species in $\W$, so
that $ m=m_A + m_B $. We denote by $\tilde \r_A, \tilde\r_B$ the
apparent densities of $A$ and $B$ respectively, namely
$$
m_A = \int_\W \tilde \r_A dv, \qquad m_B = \int_\W \tilde\r_B dv.
$$
As a consequence,
\begin{equation}\label{density}
    \r_0 = \tilde\r_A + \tilde \r_B.
\end{equation}

During phase separation each material particle cannot change its
phase, but it is only allowed to migrate from a geometrical point
to another close to it. As a consequence, the total amount of each
species in the whole domain must remain equal to the given
original amount.

In our model we consider the mixture as a single fluid obeying the
laws of conservation of mass and linear momentum of continuum
mechanics and we associate to each particle of the matter an
additional scalar function $c$, called {\it order parameter},
which allows us to distinguish one phase (fluid form) from the
other one. More precisely, we let $c=-1$ in regions filled only by
the fluid $A$ and $c = 1$ in regions where only the fluid $B$
appears.

Following the phase--field approach, we suppose that the two
immiscible fluids are not separated by a sharp interface, but we
assume that there exists a partial mixing between them in thin
layers with finite thickness  called {\it diffuse interfaces}.
Accordingly, $c$ does not take its values only in $\{-1,1\}$, but
it is allowed to vary smoothly between $-1$ and $1$ in the
interfacial regions. Moreover, if we suppose $m_A = m_B$, the
condition $c=0$ means that the fluid is in a uniform mixed state.
Such an approach traces back to van der Waals, Landau and
Ginzburg, Cahn and Hilliard (\cite{CH, landau, van}) and later it
has been developed in the theory of phase transitions (see
\cite{alt, BG, brokate, FGM_PhysicaD, gurtin, penrose_90,
penrose_93} and the references therein).

The function $c$, which we attach to each particle like as a
label, may be interpreted as the difference of local mass fraction
(concentration) of the two components, that is
$$
c \, dm = dm_A - dm_B
$$
or equivalently
$$
 c =  \frac{\tilde\r_A - \tilde \r_B}{\r_0}.
$$
In this way, it is apparent from equality \eqref{density} that
$c\in [-1,1]$. In particular, $c = -1$ (or $c=1$) wherever only
the component $A$ (or $B$) occurs.

Furthermore, the definition of $c$ guarantees that the
concentration difference of the two components is conserved in
$\W$ as the system evolves. Indeed, recalling the definition of
$\tilde\r_A$ and $\tilde\r_B$, we have
$$
\int_\W \r_0 c \,dv = \int_\W (\tilde\r_A -\tilde \r_B) dv = m_A -
m_B = {\rm constant}.
$$
In the next section, we exhibit a kinetic equation for $c$ able to
guarantee the conservation of the total concentration over the
whole domain.

\medskip

\noindent
\begin{remark}
Several authors (see \cite{gurtin,LT}, for instance) interpret the
order parameter $c$ as the local concentration of one of the
component of the binary fluid. In our notation, it means that $c$
is defined as
$$
c = \frac{\tilde\r_A}{\r_0}.
$$
As a consequence, $c\in [0,1]$.
\end{remark}

Henceforth, we denote by $t$ the time variable, $\bx$, $\bv$ the
position vector and the velocity of the particle at time $t$ in
the actual configuration, $\theta$ the absolute temperature. Also,
$\nabla$ is the gradient operator, the superposed dot is the
material derivative and $\partial_\chi$ denotes the partial
derivative with respect to the variable $\chi$. In particular,
$\partial_t$ is the partial time derivative and $\partial_j =
\partial_{{x_j}}$. Hence, for any function $g(\bx,t)$, we have
\begin{equation}\label{*}
    \dot{g} = \partial_t g + \bv \cdot \nabla g,
\end{equation}
where $\cdot$ stands for the scalar product. In addition, we use
the symbols $\nabla\cdot$ and $\Delta$ to indicate the divergence
and  the Laplacian respectively. Finally, the inner product of two
second order tensors $\bA$ and $\bB$ is defined by
$$
\bA: \bB = {\rm tr}(\bA^T\bB),
$$ where tr$\bA$ and $\bA^T$ are the trace and the transpose of a tensor $\bA$.

For reader's convenience, we briefly recall a lemma we will use in
the sequel.
\begin{lemma}\label{lemma}
For any $C^2$ function $g(\bx, t)$ the derivatives
$\dot{\overline{\nabla g}}$ and $\nabla\dot  g$ are related by the
identity
\begin{equation}\label{identity}
    \dot{\overline{\nabla g}}  =  \nabla\dot g - (\nabla \bv)^T
    \nabla g.
\end{equation}
\end{lemma}
\noindent {\it Proof.} From \eqref{*}, it follows
$$
\dot{\overline{\partial_j \vf}} = \partial_t \partial_j \vf + v_k
\,\partial_k \partial_j \vf =
\partial_j (\partial_t \vf + v_k \, \partial_k \vf) - \partial_jv_k \,\partial_k \vf,
$$
that is exactly \eqref{identity}.
\hfill$\square$

\section{Generalized Cahn--Hilliard equation}
In this section we introduce the kinetic equation for the
phase--field $c$. As we have remarked in Sect.1, since there is no
mass transfer from one phase to the other one, the mass of each
component is conserved in $\Omega$. This means that the evolution
of $c$ has to be subject to the constraint
\begin{equation}\label{int_c}
    \int_\W \rho_0\,c(\bx,t) dv = {\rm constant.}
\end{equation}

In their original papers, Cahn and Hilliard \cite{CH, C}
postulated a generalized mass diffusion equation, valid in the
entire two phase system, to describe the process of phase
separation of two components in a binary alloy under isothermal
and isochoric conditions. In particular, they assume that the
concentration of one of the two metallic components of the alloy
obeys the equation
\begin{equation}\label{CH}
\partial_t c = \div \bJ,
\end{equation}
where the local diffusion mass flux $\bJ$ satisfies the boundary
condition
\begin{equation}\label{bcJ}
    \bJ \cdot \bn |_{\partial \W} =0.
\end{equation}
In addition, $\bJ$ is assumed to be proportional to the gradient
of the generalized chemical potential $\mu$, {\it i.e.}
$$
\bJ = M(c)\nabla\mu,
$$
where $M(c)$ denotes the diffusive mobility and it is a
non--negative function eventually depending on the concentration.
The dependence of mobility on the concentration appears for the
first time in the original derivation of the Cahn--Hilliard
equation (see \cite{CH}) and later other authors considered
different expressions for $M(c)$ (see for instance \cite{Barrett,
elliott}).
The case $M(c)=0$ corresponds to a pure transport of the
components without diffusion.

Cahn and Hilliard take $\mu$ in the form
\begin{equation}\label{mu}
     \mu =  -\gamma \Delta c+ f(c),
\end{equation}
where $\gamma$ measures the width of the diffusive layer and $f$
is a double--well function, whose wells represent the two bulk
phases.

Later, this model has been applied to other contexts concerning
two phase systems constituted by different substances, for
instance air and water or oil and water. Furthermore, the
Cahn--Hilliard equation has been coupled with a kinetic equation
for the absolute temperature. For these models thermodynamically
consistency and results concerning existence, uniqueness and
long--time behaviour of the solutions have been proved (see e.g.
\cite{brokate,Gal_Mir,Mir_Sch}).

To account for fluid motion, some authors analyze the so--called
Navier--Stokes--Cahn--Hilliard system (see for instance
\cite{Gal,gurtin_PV,LT}). They substitute the partial derivative
of $c$ with respect to $t$ with the material derivative in
\eqref{CH}, {\it i.e.}
\begin{equation}\label{KE}
\rho_0 \dot{c}\,=\,\nabla \cdot [M(c)\nabla\mu]
\end{equation}
where $\mu$ is given in \eqref{mu} and they consider a modified
Navier--Stokes equation including a surface tension source term for
the coupling between $c$ and $\bv$. In such a way, equation
\eqref{KE} is composed of both a transport term, $\bv\cdot \nabla
c$, accounting for mechanical effects and due to the presence of
the material derivative, and a diffusive term at the right--hand
side modelling the chemical effects. The drawback of this model is
that when slow processes are considered, namely $\dot c \approx
\partial_t c$, the coupling between $c$ and $\bv$ disappears.

In our paper we propose a thermodynamically consistent model for
phase separation phenomena with a different coupling with the
fluid motion and including thermal effects. In particular, we
assume that $c$ satisfies equation \eqref{KE} where the chemical
potential $\mu$ is allowed to depend on the curl of the velocity.
More precisely, $\mu$ is taken in the form
\begin{equation}\label{mi}
    \mu = -\gamma\Delta c + \theta_0 F'(c) + [\theta + |\rot\bv|^2]G'(c),
\end{equation}
where $\gamma$ and $\theta_0$ are positive constants and $F, G$
are suitable function depending only on $c$ whose expressions are
given in the sequel. Accordingly, the explicit form of the
Cahn--Hillard equation is
\begin{equation}\label{CH1}
\rho_0 \dot{c}\,=\,\nabla \cdot \left\{ M(c)\nabla \left[ -\gamma
\Delta c+\theta _{0}F^{\prime }(c)+\left( \theta +\left| \nabla
\times \bv\right| ^{2}\right) G^{\prime }(c)\right] \right\}.
\end{equation}
We append to such an equation homogeneous Neumann boundary
conditions both for the difference concentration and the chemical
potential, {\it i.e.}
\begin{equation}\label{bc_mu}
    \nabla c \cdot \bn|_{\partial \W} =0,
    \qquad
    \nabla \mu \cdot \bn|_{\partial \W}=0.
\end{equation}
The first condition describes a ``contact angle'' of $\pi/2$
between the diffused interface and the boundary of the domain,
while the second one means that there is no mass flux through the
boundary and it ensures that \eqref{int_c} holds. Indeed, in view
of the transport and divergence theorems, we have the following
equalities:
$$
\frac{d}{dt}\int_{\W} \r_0 c\, dv = \int_\W \r_0 \dot c\, dv =
\int_{\partial \W} M(c)\nabla\mu \cdot \bn \, da =0.
$$

A typical choice of the functions $F,G$ is the following:
\begin{equation} \label{AlternativePotential}
F(c) = \frac{c^4}{4} - \frac{c^2}{2}, \qquad G(c) = \frac{c^2}{2},
\end{equation}
so that when
\begin{equation} \label{critico}
u= \theta +\left|\nabla \times \bv\right| ^{2}
\end{equation}
is constant, the function
$$
W(c)=\theta _{0}F(c)+ u G(c) = \theta_0
\frac{c^4}{4}+(u-\theta_0)\frac{c^2}{2}
$$
coincides with the double--well function $f$ given in \eqref{mu}.
\\
Accounting for the explicit expression of $F$ and $G$, we are able
to explain the existence of a critical value for the temperature
and the curl of the velocity. So, just to this aim, let us neglect
the quantity $\Delta(-\gamma\Delta c)$ in equation \eqref{CH1} and
suppose to fix the values of the temperature and the curl of the
velocity. Under these hypotheses, the evolution equation for $c$
reads
\begin{equation}\label{CHapprox}
    \r_0 \dot c \cong  \div [M(c) W''(c) \nabla c] = \div
    [K(c) \nabla c],
\end{equation}
where the diffusivity $K$ is defined as $K(c)=M(c)W''(c)$ and
$W''$ is given by
$$
W''(c) = 3\theta_0 c^2 + u- \theta_0.
$$
Note that since $M(c)$ is a non--negative function, the
qualitative behavior of the solution depends on the function $W$.
Precisely, when $\theta$ or $|\rot\bv|^2$ are sufficiently large
(that is $u$ is sufficiently large), the diffusion coefficient
$K(c)$ is positive since $W$ is convex and it attains a (unique)
minimum at $c=0$, {\it i.e.} the mixed phase is stable. On the
other hand, if $u <\theta_0$, then $W$ has two minima at $c=\pm
\sqrt{\frac{\theta_0 -u}{\theta_0}}$ and a local maximum at $c=0$.
This means that when $u <\theta_0$ $K(c)$ is negative in the
so--called {\it spinodal interval} $(-c_1, c_1)$ with
$$
c_1 = \sqrt{\frac{\theta_0 -u}{3\theta_0}} \ ,
$$
and it is positive when $c< -c_1$ or $c>c_1$. So the mixed phase is unstable and phase
separation occurs (see Fig. 1). The constant value $\theta_0$ can be interpreted as the critical temperature of the mixture.

Notice that equation \eqref{CHapprox} allows backward and forward
diffusion and the corresponding initial problem is classically not
well--posed from the mathematical point of view. For this reason,
in the Cahn--Hilliard equation the additional term
$\Delta(-\gamma\Delta c)$, which accounts for the interfacial
energy, appears.

\begin{figure}[h]\label{critico2}
\centering
\includegraphics{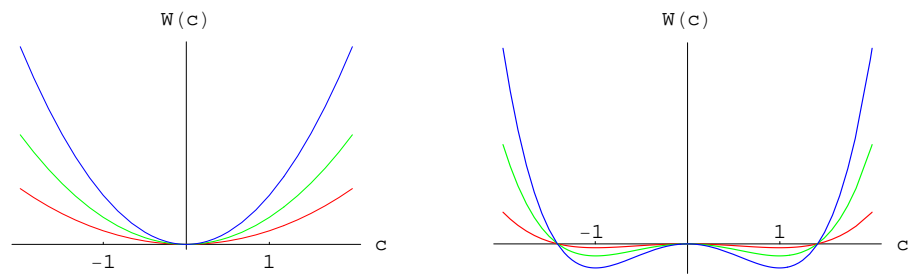}
\caption{ On the left the graphic representation of \,$W(c)$ when
$u>\theta_0$. The minimum in zero implies that, when the
temperature or the mixing velocity are sufficiently large, the
mixed phase is stable.
\newline
On the right the graphic representation of $W(c)$ when
$u<\theta_0$. In this case, we observe two minima
and the local maximum corresponding to $c\,=\,0$: the mixed
face is unstable.}
\end{figure}

%========================================================================
%========================================================================
%========================================================================
\section{Governing Equations}
The expression of the chemical potential $\mu$ involves the fluid
velocity and the temperature. Accordingly, we need to write the
kinetic equations for these variables.

By Eq. \eqref{constant}, we are modeling the fluid as an
incompressible material. Then the continuity equation provides
\begin{equation}\label{continuity}
    \div \bv = 0 \ .
\end{equation}

The linear momentum balance equation is taken in the classical
form of continuum mechanics, namely
$$
\rho_0 \dot{\bv} = \nabla \cdot \bT + \rho_0 \bb,
$$
where $\bT$ is the Cauchy stress tensor and $\bb$ is the body
force (per unit mass). We write $\bT$ as the sum of three second
order tensors, {\it i.e.}
$$
\bT= \bT^v +\hat \bT + \bS.
$$
The first term is related to the classical Cauchy stress tensor in
the Navier--Stokes equation, that is
$$
\bT^v = -p \textbf{1} +\nu(c) [\nabla \bv + (\nabla\bv)^T] = -p
\textbf{1} + 2\nu(c) \bD,
$$
where $p$ is the pressure (which is a priori unknown since we have
supposed the fluid incompressible), $\textbf{1}$ is the second
order identity tensor, $\nu$ is the viscosity of the fluid and
$\bD$ is the symmetrical part of the gradient of velocity. We
stress that $\nu$ depends on the concentration $c$, so that when
$c=1$ (or $c=-1$) $\nu$ coincides with the viscosity of the fluid
$A$ (or $B$). In the next section, we will prove that $\nu(c)>0$
as a consequence of the second law of thermodynamics.

We add to the usual (symmetric) tensor $\bT^v$ the extra reactive
stress $\hat \bT$ associated with the presence of concentration
gradient which models the capillary forces due to surface tension
(Ericksen's stress, see \cite{ericksen}), {\it i.e.}
$$
\hat \bT = -\gamma \r_0 \nabla c\otimes \nabla c,
$$
where the parameter $\gamma$ is assumed to be positive and it is
related to the thickness of the interfacial region. This term
occurs even on other papers concerning with the
Navier--Stokes--Cahn--Hilliard equation (see \cite{gurtin,LT}).

Finally, we introduce the skew tensor $\bS$ whose components are
defined as
$$
S_{ij} = \e_{ijk}\, \r_0\dot G(c) (\rot\bv)_k,
$$
where $\e_{ijk}$ denotes the Levi--Civita symbol and summation is
implied by index repetition. This term makes the tensor $\bT$
non--symmetric, consistent with the presence of internal structure
due to the mixture (see \cite{cosserat}). However, $\bS$
disappears when $\dot G =0$, that is in the bulk phases.

By evaluating the $i-$th component of the divergence of $\bS$, we
obtain
$$
(\div\bS)_i = \partial_j S_{ij} = \partial_j [\e_{ijk} \, \r_0
\dot G(c) (\rot\bv)_k] =  \e_{ijk} \partial_j [\r_0 \dot G(c)
(\rot\bv)_k]
$$
namely,
$$
\div\bS = \r_0 \rot[\dot G(c)\rot\bv].
$$
Accordingly, the velocity $\bv$ satisfies the modified
Navier--Stokes equation
\begin{equation}
\label{v}
    \rho_0 \dot \bv = -\nabla p + \div\{2\nu(c) \bD \}  - \gamma \r_0 \div(\nabla c \otimes \nabla c) + \r_0 \rot[\dot G(c)\rot\bv]
    + \rho_0 \bb.
\end{equation}
To this equation we associate the usual no--slip boundary
condition:
$$
\bv|_{\partial\Omega} = {\bf 0}.
$$

In order to obtain the kinetic equation for the temperature, let us consider the first law of thermodynamics as in \cite{fremond} or \cite{fried-gurtin}
\begin{equation}\label{Ilaw}
    \r_0\dot E = \cP^i_m + \cP^i_c + \rho_0 h,
\end{equation}
where $E$ is the total energy, $\cP^i_m, \cP^i_c$ are respectively the internal mechanical and chemical power whose expressions are given in the sequel, and $h$ stands for the rate at which the heat is absorbed by the material. Denoting with $T =
\frac{1}{2} \bv^2$ the kinetic energy and $e$ the internal energy,
which we suppose function of the state $\sigma = (\theta,c,\nabla
c)$ of the system, we write $E= T+ e$.

By multiplying equation \eqref{v} by $\bv$ and accounting for
\eqref{continuity}, we obtain the power balance related to the
velocity $\bv$, that is \bea \cP^i_m = \cP^e_m, \eea with
\begin{eqnarray}
\label{mech} \cP^i_m &=& \frac12 \rho_0 \frac{d}{dt}\bv^2
+\nu(c)|\nabla\bv|^2 + \nu(c)(\nabla\bv)^T : \nabla\bv - \gamma
\r_0
 (\nabla c\otimes \nabla c) : \nabla \bv
\\ \nonumber
&& - \r_0 \dot G(c) |\rot\bv|^2 ,
\\
\cP^e_m &=& \div [-p\bv + \nu(c) \bD \bv  - \gamma \r_0 (\nabla
c\otimes \nabla c)\bv + \r_0 \dot G(c) (\rot\bv) \times \bv] +
\r_0 \bb\cdot \bv.
\end{eqnarray}

Similarly, multiplying equation \eqref{mi} by $\r_0 \dot c$ and
taking  \eqref{KE} into account, we obtain the power balance
related to the concentration $c$, that is \bea \cP^i_c = \cP^e_c,
\eea where
\begin{eqnarray}
\label{chem} \cP^i_c &=& \rho_0 \theta_0 \dot F(c) + \rho_0
\dot{G}(c)[\theta + |\rot\bv|^2] + \rho_0 \gamma \nabla c\cdot
\nabla \dot c  + M(c) |\nabla\mu|^2,
\\
\cP^e_c &=& \div [\rho_0 \gamma  \dot c \nabla c + M(c)\mu
\nabla\mu].
\end{eqnarray}
\newline
By means of Lemma \ref{lemma} that applied to the concentration
$c$ yields \begin{equation} \label{DerivLemma}
\dot{\overline{\nabla c}} \cdot \nabla c = [\nabla\dot c - (\nabla
\bv)^T \nabla c] \cdot \nabla c = \nabla\dot c\cdot \nabla c  -
(\nabla c \otimes \nabla c):\nabla\bv,
\end{equation}
 and summing up
$\cP^i_m$ and $\cP^i_c$, we obtain
\begin{equation}\label{sommaP}
\cP^i_m + \cP^i_c = \rho_0 \disp\frac{d}{dt}\left[\frac12 \bv^2
+ \theta_0 F(c) + \frac12 \g|\nabla c|^2 \right] + \nu(c)
|\nabla\bv|^2
 + \nu(c) (\nabla \bv)^T : \nabla \bv + \rho_0 \theta\dot{G}(c)
+ M(c) |\nabla\mu|^2.
\end{equation}
We stress that the fourth term in the rhs of equation \eqref{mech}
is exactly $\hat \bT : \nabla \textbf{v}$, hence by
\eqref{DerivLemma}, the contribute of $\hat \bT$ to the internal
power is enclosed into $\rho_0 \frac{d}{dt}(\frac{1}{2} \gamma
|\nabla c|^2)$.

Moreover, \eqref{sommaP} suggests to define the internal energy
$e$ as
\begin{equation}\label{internal_energy}
    e(\sigma) = e_0(\theta)  + \theta_0 F(c)
    + \frac12 \g|\nabla c|^2,
\end{equation}
where $e_0$ is a suitable function depending only on the
temperature. Then, the total energy $E$ is given by
$$
E  = T + e = \frac{1}{2}\bv^2+ e_0(\theta)  + \theta_0 F(c)
    + \frac12 \g|\nabla c|^2.
$$
As a consequence, a comparison with \eqref{Ilaw} yields
\begin{equation}\label{h}
    \rho_0 h = \rho_0 \dot \theta  e_0'(\theta) - \nu(c)|\nabla \bv|^2 - \nu(c)(\nabla\bv)^T : \nabla\bv
     - \rho_0 \theta \dot G(c) - M(c) |\nabla\mu|^2.
\end{equation}

In our model the Fourier theory of heat conduction will not be
modified. Accordingly, the constitutive equation relating the heat
flux $\bq$ to the gradient of the temperature, assumes the
classical form
\begin{equation}\label{FourLaw}
\bq = - \kappa(\theta) \nabla \theta,
\end{equation}
where $\kappa(\theta)>0$ denotes the thermal conductivity and it
depends on the absolute temperature.
\newline
As well known (see e.g. \cite{fremond}), the thermal balance law is expressed by the
following equation
$$
\rho_0 h = -\div \bq + \rho_0 r.
$$
Comparing with \eqref{h}, we have
$$
\rho_0 \dot \theta e'_0(\theta) - \nu(c)|\nabla \bv|^2 -
\nu(c)(\nabla\bv)^T : \nabla\bv
 - \rho_0 \theta \dot G(c) - M(c) |\nabla\mu|^2 = -\div \bq + \rho_0 r.
$$

Finally, collecting the equations of motion we write the system of
equations:
\begin{equation}
\label{system}
\begin{array}{lll}
  \div \bv&=&0
  \\
  \noalign{\medskip}
\rho_0 \dot \bv &=& -\nabla p + \div[\nu(c) \bD]  - \gamma \rho_0
\div(\nabla c \otimes \nabla c) + \rho_0 \rot[\dot G(c) \rot\bv] +
\rho_0\bb
\\
\noalign{\medskip} \rho_0 \dot c &=& \,\nabla \cdot \left[
M(c)\nabla \left( -\gamma \Delta c+\theta _{0}F^{\prime
}(c)+\left[ \theta +\left| \nabla \times \bv\right| ^{2}\right]
G^{\prime }(c)\right) \right]
\\
\noalign{\medskip} \rho_0 \dot \theta  e_0'(\theta)
 &=& \nu(c)|\nabla \bv|^2 + \nu(c)(\nabla\bv)^T : \nabla\bv + \rho_0 \theta \dot G(c) +  M(c) |\nabla\mu|^2  -\div \bq + \rho_0 r
 \end{array}
\end{equation}
and we associate to \eqref{system} the boundary conditions \bea
\nabla c \cdot \bn |_{\partial \Omega} &=&0, \qquad \nabla \mu
\cdot \bn |_{\partial \Omega} =0,
\\
\bv |_{\partial \Omega} &=& {\bf 0}, \qquad\, \nabla \theta \cdot
\bn |_{\partial \Omega} =0. \eea and the initial data \bea
c(\bx,0) = c_0(\bx), \qquad \bv(\bx,0) = \bv_0(\bx), \qquad
\theta(\bx,0) = \theta_0(\bx). \eea

%===============================================================
%===============================================================
%===============================================================
\section{Thermodynamics}
In this section we show that our model is consistent with the
second law of thermodynamics written in the Clausius--Duhem form.

\medskip
\noindent \textit{Second law of thermodynamics.} There exists a
function $\eta$, called entropy function, such that
\begin{equation}\label{EntrIneq}
\rho_0 \dot\eta \geq -\div \left(\frac{\bq}{\theta} \right) +
\frac{\rho_0 r}{\theta},
\end{equation}
where $\bq$ is the heat flux vector and $r$ is the external heat
supply density.

We introduce the Helmholtz free energy density $\psi$ defined as
$$
\psi(\sigma) = e(\sigma) - \theta \eta(\sigma).
$$
Now we are in a position to prove the following result.
\begin{theorem}
The functions $\bq$, $\psi$ and $\eta$ are compatible with the
second law of thermodynamics if and only if the viscosity $\nu$,
the mobility $M$ and the thermal conductivity $\kappa$ are
non--negative functions and the free energy $\psi$ satisfies the
following conditions:
\begin{equation}\label{therm_restr}
    \partial_\t \psi = -\eta, \qquad \partial_c \psi = \t_0 F'(c) + \theta G'(c),
    \qquad
    \partial_{\nabla c} \psi = \gamma \nabla c.
\end{equation}
\end{theorem}

\medskip
\noindent \noindent {\it Proof.} In order to obtain compatibility
with thermodynamics, we have to prove that system \eqref{system}
with constitutive equations \eqref{h}, \eqref{FourLaw} agrees with
inequality \eqref{EntrIneq}, which, by means of the thermal
balance law, becomes
\begin{equation}\label{IIlaw}
    \rho_0 \dot\eta \theta \geq \frac{1}{\theta} \nabla\theta \cdot \bq + \rho_0 h.
\end{equation}
Moreover, by the free energy density $\psi = e - \theta \eta$,
\eqref{IIlaw} can be written as
$$
\rho_0 \dot\psi  - \rho_0\dot e + \rho_0\dot \theta \eta + \rho_0
h + \frac{1}{\theta} \nabla\theta \cdot \bq \leq 0.
$$
In view of \eqref{internal_energy}-\eqref{h} we have
\begin{eqnarray}\label{CD_psi}
\nonumber && \rho_0\left(\partial_\t \psi +\eta \right)\dot\t
+\rho_0\left[\partial_c \psi - \t_0 F'(c) -\theta G'(c)
\right]\dot c +\rho_0\left(\partial_{\nabla c} \psi - \gamma
\nabla c \right) \cdot \dot{\overline{\nabla c}}
\\
\label{CD} &&
 - \nu(c)|\nabla \bv|^2 - \nu(c)(\nabla\bv)^T : \nabla\bv  - M(c) |\nabla\mu|^2 + \frac{1}{\theta} \nabla\theta\cdot \bq \leq 0.
\end{eqnarray}
From definition \eqref{mi}, it follows that
$$
\nabla \mu = -\gamma \nabla(\Delta c) + \nabla\{\theta_0 F'(c) +
[\theta + |\rot\bv|^2]G'(c)\}.
$$
Since $\nabla(\Delta c)$ may be chosen arbitrarily, $\nabla\mu$
may be chosen arbitrarily too. Accordingly, by standard arguments,
we deduce the following conditions: \bea
\partial_\t \psi = -\eta, \qquad \partial_c \psi = \t_0 F'(c) + \theta G'(c),
\qquad
\partial_{\nabla c} \psi = \gamma \nabla c
\eea and, in view of \eqref{FourLaw}, we write inequality
\eqref{CD_psi} in the form
$$
- \nu(c)|\nabla \bv|^2 - \nu(c)(\nabla\bv)^T : \nabla\bv  - M(c)
|\nabla\mu|^2 - \frac{\kappa(\theta)}{\theta} |\nabla\theta|^2\leq
0.
$$
In addition, relation
$$
(\nabla\bv)^T : \nabla\bv = \rm{tr}(|\nabla\bv|^2) \geq 0,
$$
allows us to conclude that the function $\nu(c)$, $M(c)$ and
$\kappa(\theta)$ are non--negative. \hfill$\square$

\medskip

From \eqref{therm_restr} it follows that the free energy density
$\psi$ and the entropy $\eta$ are given, up to a constant, as
\begin{eqnarray}
\label{free_energy} \psi &=&  \theta_0 F(c) + \theta G(c) +
\frac{\gamma}{2}|\nabla c|^2 + \psi_0(\theta),
\\
\label{entropy} \eta &=& - \partial_{\theta} \psi= -G(c) -
\psi'_0(\theta),
\end{eqnarray}
where $\psi_0$ is a suitable function (depending only on $\theta$)
which ensures the validity of the condition $\psi = e- \eta\theta
= e + \partial_\theta \psi\theta$. A substitution of
\eqref{internal_energy} and \eqref{free_energy} leads to the
equality
$$
\psi_0(\theta)= e_0(\t) + \psi_0'(\theta) \theta.
$$
Thus, $\psi_0$ is given by
$$
\psi_0 = \mathcal{C} \t - \t \int\frac{e_0(\theta)}{\theta^2}
d\theta,
$$
with $\mathcal{C}>0$ and
$$
\eta = -G(c) -\mathcal{C} + \int\frac{e_0(\theta)}{\theta^2}
d\theta + \frac{e_0(\theta)}{\theta}.
$$
In particular, if we let $e_0 = \mathcal{C} \theta$, where
$\mathcal{C}$ denotes the specific heat, we recover the standard
form of $\psi_0$ and $\eta$, {\it i.e.}
$$
\psi_0 = \mathcal{C} \theta (1-\ln\theta), \qquad \eta = -G(c)
-\mathcal{C} \ln\theta.
$$

%%%%%%%%%%%%%%%%%%%%%%%%%%%%%%%%%%%%%%%%%%%%%%%%%%%%%
%%%%%%%%%%%%%%%%%%%%%%%%%%%%%%%%%%%%%%%%%%%%%%%%%%%%%
%%%%%%%%%%%%%%%%%%%%%%%%%%%%%%%%%%%%%%%%%%%%%%%%%%%%%

\ack The authors have been partially supported by G.N.F.M. --
I.N.D.A.M. through the project for young researchers ``Mathematical models for phase transitions in special materials''.

\end{document}